\documentclass[preprint]{revtex4-1}
\usepackage{pifont}
\usepackage{graphicx,colordvi}
\usepackage{amsmath,amssymb,amsthm}
\usepackage[colorlinks=true,urlcolor=blue,linkcolor=blue,citecolor=red]{hyperref}
\usepackage{mathrsfs}

\newcommand{\ini}{\begin{equation}}
\newcommand{\fin}{\end{equation}}
\newcommand{\inia}{\begin{eqnarray}}
\newcommand{\fina}{\end{eqnarray}}

\begin{document}

\title{\bf  Including gauge symmetry in the localization mechanism of massive vector fields}
\author{Rommel Guerrero  and R. Omar Rodriguez}
\address{Departamento de F\'isica, Universidad Centroccidental Lisandro Alvarado, Barquisimeto,Venezuela}

\begin{abstract}
On the four-dimensional sector of an AdS$_5$ warped geometry the standard electromagnetic interaction can be  simulated by massive vector fields via the Ghoroku - Nakamura localization mechanism.  We incorporate gauge symmetry to this theory by finding  the required interaction terms between  the vector bosons and the gravitational field of the scenario.  The four-dimensional effective theory defined by a Maxwell term and  a  tower of Stueckelberg fields is obtained after expanding the vector fields on a massive eigenstates basis where the zero mode is uncoupled from the rest of the spectrum. The corrections generated by the massive gauge fields set to the electrostatic potential are also calculated.


\vspace{0.5 cm} 
\end{abstract}

\maketitle

\section{Introduction}
In the theories with extended extra dimensions  our Universe is conceived as a  hypersurface inside a five-dimensional Anti de Sitter (AdS$_5$) spacetime  \cite{Randall:1999vf}.  Often, these scenarios are generated by a scalar field in self-interaction whose energy density is in correspondence with a transition region between two AdS$_5$ vacua. It is expected that bulk physical fields exhibit a standard behaviour on the four-dimensional sector of the theory; however, in some cases, the gravitational field prevents  this from happening. In particular, in absence of a suitable coupling term, it is not possible confining bulk gauge fields on the boundary that connects the AdS$_5$ spaces  \cite{Bajc:1999mh}. 

The vector field localization is a subject that has been discussed, in our opinion,  from two different approaches: considering corrections to the kinematics of the bosons \cite{Dvali:2000rx} or assuming no-convetional interaction terms between the vectors and the gravitational field \cite{Kehagias:2000au,Ghoroku:2001zu,Batell:2006dp}.  Here, we are interested in the second approach, specifically in the localization mechanism  reported  by Ghoroku and Nakamura in  \cite{Ghoroku:2001zu}, where the dynamics of the vector fields on an AdS$_5$ warped geometry \cite{Randall:1999vf} is determined by
\begin{equation}\label{RSmetric}
ds^2 = \left(1+\alpha|z|\right)^{-2}\left(\eta_{\mu\nu}dx^\mu dx^\nu+dz^2\right)
\end{equation}
and
\begin{equation}\label{LGN}
\frac{{\mathscr L}}{\sqrt{-g}}=-\frac{1}{4} F_{ab}F^{ab}- \frac{1}{2} \left( m_5^2-m _4^2 \delta(z)\right) A_a A^a-{\cal Q}^2 A_a J^a. 
\end{equation}
Latin and Greek indices for five- and four-dimensions respectively; $z$ reserved to the extra coordinate.

From (\ref{LGN}) it can be deduced that the four-dimensional effective propagator on the transition region is given by 
\begin{eqnarray}\label{propagator}
G^T_{\alpha\beta}(p)=-\frac{{\cal Q}^2}{\pi^2}\left(\eta_{\alpha\beta}-\frac{p_\alpha p_\beta}{p^2}\right)\left[\frac{\alpha(\nu-1)}{p^2}-\frac{i}{2}\frac{H^{(1)}_{\nu-2}(ip/\alpha)}{H^{(1)}_{\nu-1}(ip/\alpha)}\ \frac{1}{p}\right]
\end{eqnarray}
where $T$ means transverse to the four-momentum $p^\beta$ and $\nu^2=m_5^2/\alpha^2+1$.  

Notice that the mechanism relaxes the gauge symmetry and appeals to the interaction between bulk massive vector fields and the topological defect at $z=0$. However,  the infrared behaviour of (\ref{propagator}), $p\ll\alpha$, resembles to the standard electromagnetic one (tensorial structure omitted),
  \begin{eqnarray}\label{Gghoroku}
 G^{T}(p,0,0)\sim\frac{\alpha(\nu-1)}{p^2}-\frac{1}{4\alpha(\nu-2)}+\frac{\pi i^{2\nu+1}}{2^{2(\nu-1)}}\frac{(1+i \cot\pi\nu)}{\Gamma(\nu-1)^2\alpha^{2\nu-3}}p^{2(\nu-2)}.  
 \end{eqnarray}
 On the other hand, in the ultraviolet regime, $p\gg\alpha$, we have that 
$H^{(1)}_{\nu-2}(ip/\alpha)/H^{(1)}_{\nu-1}(ip/\alpha)\sim i$ and as a result 
\begin{equation}G^T(p,0,0)\sim 1/p.\end{equation} 

The model (\ref{LGN}) is a propose  to recover standard electromagnetism on the four-dimensional sector of (\ref{RSmetric}), however the absence of gauge symmetry weakens the result (\ref{Gghoroku}) because, after all,  the mediator of interaction is a massive vector field. The Lagrangian (\ref{LGN}) should be obtained after fix gauge on an $U(1)$-invariant theory; approaches in this direction could be varied, our proposal  is to generalize the mechanism (\ref{LGN}) to scenarios determined by a scalar field in self-interaction and incorporate the gauge symmetry by means of appropriate interaction terms between the vector bosons and  gravitational field of the model.  

\section{Including gauge symmetry}
Let us to start considering the gravitational background   
\begin{equation}\label{LKG}
 \frac{{\mathscr L}_g}{\sqrt{-g}}=\frac{1}{2}R-\frac{1}{2} \nabla^a\phi\nabla_a\phi-V(\phi)
 \end{equation}
where the scalar field interpolates between the minima of the self-interaction potential  \cite{Gremm:1999pj, DeWolfe:1999cp,CastilloFelisola:2004eg}. In particular, for the coordinates
\begin{equation}\label{Metric}
 ds^2=e^{2a(z)}\left(\eta_{\mu\nu}dx^\mu dx^\nu+dz^2\right)
 \end{equation}
 we have 
 \begin{equation}\label{scalarfield}
 \phi^{\prime2}=3\left[a^{\prime 2}-a^{\prime\prime}\right],
 \end{equation}
 \begin{equation}\label{selfpotential}
 V(\phi)=-\frac{3}{2}\left[3a^{\prime 2}+a^{\prime\prime}\right]e^{-2a},
 \end{equation}
where $a^\prime=\partial_z a$ and $a^{\prime\prime}=\partial_z^2a$.

On the scenarios generated from (\ref{LKG}, \ref{Metric}) we propose a no-conventional dynamics for the electromagnetic field  
\begin{eqnarray}\label{SAInv}
\frac{{\mathscr L}_A}{\sqrt{-g}}=-\frac{1}{4}{F}_{ab}F^{ab}-\left(\kappa-\frac{1}{2}\right)g^{a[c}  g^{d]b} a^\prime \delta^z_c\left[ -2 \partial_a A_b+\left(\kappa-\frac{1}{2}\right)a^\prime \delta^z_a A_b\right] A_d ;
\end{eqnarray}
which is invariant under the gauge transformation 
\ini\label{LTA}
\delta {A}_b=\partial_b{\chi}-\left(\kappa-\frac{1}{2}\right) a^\prime \delta^z_b{\chi}  .
\fin

Clearly $\kappa$ is the coupling parameter; namely, for $\kappa=1/2$ the interaction terms disappear and the Lagrangian evidently is $U(1)$-invariant.  Unfortunately, in this case, it is not feasible to determinate a four-dimensional effective theory in correspondence with the standard electromagnetic interaction \cite{Bajc:1999mh}. For example, in the gauge $A_z=0$ and under the factorization $A_\mu(x,z)=a_\mu(x)\varphi(z)$ we get
\begin{eqnarray}
-\frac{1}{4}\int dx^4 dz \sqrt{-g}F^{ab}F_{ab}=&-&\frac{1}{4}\int dx^4 \eta^{\mu\alpha}\eta^{\nu\beta}f_{\alpha\beta}f_{\mu\nu}\int_{-\infty}^\infty dy \varphi^2\nonumber\\&-&\frac{1}{2}\int dx^4 \eta^{\mu\alpha} a_\alpha a_\mu\int_{-\infty}^\infty dy (\partial_y\varphi)^2,\label{U(1)}
\end{eqnarray}
with $f_{\alpha\beta}=\partial_\alpha a_\beta-\partial_\beta a_\alpha$ and  $dy=e^a dz$, which diverges for $\varphi=$ctte. It can be verified that  it is not possible to find  normalizable solutions for the equation of motion of the vector fields. 

For $\kappa>1/2$, no conventional interaction terms are incorporated to (\ref{SAInv}) and make visible the gauge symmetry  by simple inspection it is not easy; it shows obvious after transforming each of the terms of (\ref{SAInv}):
\begin{eqnarray}
&&\delta\left(-\frac{1}{4} {F}_{ab}F^{ab}\right)= 2\left(\kappa-\frac{1}{2}\right) {a^\prime \delta^z_d g^{a[c}g^{d]b} \partial_a A_b \partial_c\chi};\\\nonumber\\
&&\delta\left(g^{a[c}  g^{d]b} a^\prime \delta^z_c A_d \partial_a A_b\right)= a^\prime \delta^z_{c}  g^{a{[c}}  g^{{d]}b}\left[ {\partial_a A_b\partial_{d}\chi}-\left(\kappa-\frac{1}{2}\right){a^\prime \delta^z_bA_d\partial_a  \chi}\right];\\\nonumber\\
&&\delta\left(- a^{\prime 2}   \delta^z_a  \delta^z_c g^{a[c}  g^{d]b} A_d A_b\right)=2 {a^{\prime 2}   \delta^z_b  \delta^z_c g^{a[c}  g^{d]b} A_d \partial_a{\chi}}.
\end{eqnarray}
Combining these three terms we obtain $\delta({\mathscr L_A})=0$.

Now, it is convenient to rewrite (\ref{SAInv}) as follows 
\begin{eqnarray}\label{Full}
\frac{{\mathscr L}_A}{\sqrt{-g}} =\frac{\tilde{{\mathscr L}}_A}{\sqrt{-g}}&-&\frac{2}{3}V(\kappa,z) e^{2a} \delta^z_b \delta^z_c  A^b A^c\nonumber\\&-&\left(\kappa-\frac{1}{2}\right)g^{a[c}g^{d] b}\left[\left(a^{\prime\prime}+a^{\prime 2}\right) \delta^z_a A_d+2 a^\prime\partial_a A_d \right] \delta^z_c A_b,
\end{eqnarray}
where 
\begin{equation}\label{LA1}
\frac{\tilde{{\mathscr L}}_A}{\sqrt{-g}}= -\frac{1}{4} F_{ab}F^{ab}+ \frac{2}{3}V(\kappa,z)  A_a A^a
\end{equation}
and 
\ini
V(\kappa,z)=-\frac{3}{4}\left(\kappa-\frac{1}{2}\right)\left[\left(\kappa+\frac{1}{2}\right) a^{\prime 2}+a^{\prime\prime}\right] e^{-2a}.
\fin

The rearrangement of terms highlights to $\tilde{{\mathscr L}}_A/\sqrt{-g}$ as a particular portion  of (\ref{Full}), which, in isolation, can be considered as a vector field theory where the gauge symmetry is not present. This sector corresponds to a generalization of (\ref{LGN}) to geometries defined by self-gravitating domain walls; notice that for $\kappa=5/2$, $V(5/2,z)$ coincides with the scalar potential (\ref{selfpotential}). For the static AdS$_5$ case, the geometrical configurations generated from (\ref{LKG}) are  regularized representations of Randall-Sundrum scenario II \cite{Randall:1999vf} and it is expected that in the zero thickness  limit (\ref{LA1}) converges to (\ref{LGN}); in fact 
\begin{eqnarray}
\frac{\tilde{{\mathscr L}}_A}{\sqrt{-g}}\rightarrow-\frac{1}{4} F_{ab}F^{ab}-\frac{1}{2}\left[\frac{1}{6}\left(\kappa-\frac{1}{2}\right)\left(\kappa+\frac{3}{2}\right)|\Lambda|-\frac{1}{3}\left(\kappa-\frac{1}{2}\right)\tau\delta(z)\right]A_a A^a
\end{eqnarray}
where five- and four-dimensional masses of (\ref{LGN}) are determined, respectively,  by  the bulk cosmological constant $\Lambda=-6\alpha^2$ and the thin wall tension $\tau=6\alpha$ of the warped geometry (\ref{RSmetric}). Therefore,  the required extension to include gauge symmetry in the model (\ref{LGN}) is the full Lagrangian (\ref{Full}). 

\subsection{On the wall}

After having established a gauge field theory on an AdS$_5$ domain wall geometry, the four-dimensional effective theory must be compared with the standard electromagnetic interaction.  To this end, first, in (\ref{Full}) we will redefine the vector field as follows
\begin{equation}
A_b=e^{-a/2}A_b,
\end{equation} 
and then will appeal to a pair of regulatory walls located at $\pm z_r$ in order to expand the fields in a discrete basis; the original  scenario should be obtained after taking the limit  $z_r\rightarrow\infty$ (see \cite{Callin:2004py} for full details about the method of the regulatory branes). Thus, for the first four components of $A_b$ we have
\ini\label{Amupsi}
A_\mu(x,z)=a_\mu(x)\psi_0(z)+\sum_{n\neq 0} a_\mu^n(x)\psi_n(z) ,
\fin
while for the last component  
\ini\label{Azphi}
A_z(x,z)=\sum_{n\neq 0} a_z^n (x)\varphi_n(z),
\fin
where $\psi_n(z)$ and $\varphi_n(z)$, respectively,  satisfy 
\ini\label{QQ+psi}
\boldsymbol{Q} \boldsymbol{Q}^+\psi_n=m_n^2\psi_n,\quad  \boldsymbol{Q}^+\psi_n{\Big |}_{\pm z_r}=0
\fin
and
\ini\label{Q+Qpsi}
\boldsymbol{Q}^+ \boldsymbol{Q}\varphi_n=m_n^2\varphi_n,\quad \varphi_n=\frac{\boldsymbol{Q}^+\psi_n}{m_n}
\fin
with $\boldsymbol{Q}$ and  $\boldsymbol{Q}^+$ given by 
\ini
\boldsymbol{Q} =\partial_z+\kappa a^{\prime},\quad \boldsymbol{Q}^+ =-\partial_z+\kappa a^{\prime}.
\fin
Additionally the following orthogonality relations will be considered 
\ini\label{Base}
\int_{-z_r}^{z_r} dz\  \psi_n \psi_p=\delta_{np},\quad \int_{-z_r}^{z_r} dz\  \varphi_n \varphi_p=\delta_{np}.
\fin
 
Notice that the eigenvalues problem (\ref{QQ+psi}) is defined for $m^2_n\geq 0$ with a first   eigenstate determined by 
 \begin{equation}\label{psi0}
{\psi_0(z)=N_0\ e^{\kappa a(z)}};
 \end{equation}
in contrast with the problem (\ref{Q+Qpsi}) whose eigenstates are defined strictly for $m^2_n> 0$.

By replacing (\ref{Amupsi}) and (\ref{Azphi}) into (\ref{Full}) and integrating with respect to extra coordinate, the four-dimensional effective theory is obtained 
\ini\label{L4}
{\mathscr L}^{(4)}_A= -\frac{1}{4} f_{\alpha\beta}^2+\sum_n\left[-\frac{1}{4} \left(f_{\alpha\beta}^{n }\right)^2-\frac{1}{2} \left(m_n a_\mu^{n}+\partial_\mu a_z^n\right)^2\right],
\fin
where 
\begin{eqnarray}
f_{\alpha\beta}&=&\partial_\alpha a_\beta(x)-\partial_\beta a_\alpha(x),\\
f^n_{\alpha\beta}&=&\partial_\alpha a^n_\beta(x)-\partial_\beta a^n_\alpha(x),
\end{eqnarray}
and
\ini
\delta a_\mu=\partial_\mu\chi_0 ,\qquad \delta a_\mu^n=\partial_\mu\chi_n,\quad  \delta a_z^n=-m_n \chi_n.
\fin

Remarkably, the effective  theory is determined by a Maxwell field and a tower of Stueckelberg fields; i.e. at high energies the massive gauge fields intervene in the the electromagnetic interaction.  If   the corrections  generated by the massive vectors are not significant, the standard electromagnetic interaction can be recovered on the four-dimensional sector of (\ref{Full}). Therefore, the order of the corrections in the electrostatic potential must be calculated.

Next, in (\ref{Full}) we will introduce  a source for the electromagnetic radiation $-{\cal Q}^2A_bJ^b(x,z)$ and a gauge-fixing term $-(\lambda\delta^b_zA_b)^2/2\zeta$ with  $\zeta\rightarrow 0$ corresponding to the unitary gauge  $A_z=0$. As a starting point to determine the order of corrections we will consider the potential, in the momentum space, associated with two static sources 
\begin{eqnarray}
{\mathscr U}_J=\frac{(\sqrt{2\pi})^3}{2}\int d^3p\int dz\sqrt{-g(z)}\int d\xi\sqrt{-g(\xi)}\tilde{J}_1^a(\vec{p},z)\tilde{G}_{ab}(\vec{p},z,\xi)\tilde{J}_2^b(-\vec{p},\xi).
\end{eqnarray}
where $\tilde{G}_{ab}(\vec{p},z,\xi)$ is the propagator of $A_b$. In particular, for current densities strongly localized on the four-space, $\tilde{J}^a(\vec{p},z)=\delta^a_\mu\tilde{j}^\mu(\vec{p})\delta(z)$, we get
 \begin{equation}
{\mathscr U}_j=\frac{(\sqrt{2\pi})^3}{2}\int d^3p\tilde{j}^\alpha_1(\vec{p})\tilde{G}_{\alpha\beta}(\vec{p},0,0)\tilde{j}^\beta_2(-\vec{p}).\label{Uj}
 \end{equation}

The $(\alpha, z)$ and $(\alpha, \beta)$ components of propagator are determined by a coupled system of equations; however, $\tilde{G}_{\alpha z}\rightarrow 0$ for $\zeta\rightarrow 0$ and the equations system is uncoupled in such a way that four-dimensional components are given by
\begin{eqnarray}
-\left[\left(\eta^{\nu\alpha}-\frac{\bar{p}^\nu\bar{p}^\alpha}{\bar{p}^2}\right)\bar{p}^2-\frac{4}{3}V(\kappa,z)e^{2a}\eta^{\nu\alpha}-e^{-a}\eta^{\nu\alpha}\partial_z\left(e^a\partial_z\right)\right]\tilde{G}_{\alpha\beta}=\frac{{\cal Q}^2e^{-a}\delta^\nu_\beta}{(2\pi)^2}\delta(z-\xi)
\end{eqnarray}
where $\bar{p}^\alpha=\eta^{\alpha\beta} p_\beta$.

Writing $\tilde{G}_{\alpha\beta}$ in the form 
\begin{equation}\label{G}
\tilde{G}_{\alpha\beta}=\left(\eta_{\alpha\beta}-\frac {p_\alpha p_\beta}{\bar{p}^2}\right)G_1+\frac{p_\alpha p_\beta}{\bar{p}^2}G_2
\end{equation}
it can be seen that  $G_1$ and $G_2$ satisfy
 \begin{equation}
\left[e^{-a}\partial_z\left(e^{a}\partial_z\right)-\bar{p}^2+\frac{4}{3}V(\kappa,z) e^{2a} \right]G_1=\frac{{\cal Q}^2e^{-a}}{(2\pi)^2}\delta(z-\xi)\label{G1}
\end{equation}
and
\begin{equation}
\left[e^{-a}\partial_z\left(e^{a}\partial_z\right)+\frac{4}{3}V(\kappa,z) e^{2a}\right] G_2=\frac{{\cal Q}^2e^{-a}}{(2\pi)^2} \delta(z-\xi)\label{G2}
\end{equation}
respectively. Note that $G_2$ is independent of the momentum.

Now, expanding  $G_1$ in the eigenfunctions tower of the  problem (\ref{QQ+psi}), 
\begin{equation}\label{G1m}
G_1(p, z, \xi)=-\frac{{\cal Q}^2e^{-\left[a(\xi)+a(z)\right]/2}}{(2\pi)^2}\sum_n \frac{\psi^*_n(\xi)\psi_n(z)}{p^2+m_n^2} ,
\end{equation}
and considering that the eigenvalues $m_n$ are approximately quantized in units of $\pi/z_r$ for $z_r\rightarrow\infty$, the electrostatic potential (\ref{Uj}), in the coordinates space, between two charged particles $q_1$ and $q_2$; i.e., $j^\mu_i(\vec{x})=q_i\delta(\vec{x}-\vec{x}_i)\delta_0^\mu$, is given by 
\begin{eqnarray}\label{Vr}
{\mathscr U}(r)=\frac{{\cal Q}^2|\psi_0(0)|^2}{4(\sqrt{2\pi})^5}\frac{q_1q_2}{r}\left(1+\frac{1}{\pi|\psi_0(0)|^2}\int_0^\infty z_r|\psi_m(0)|^2\ e^{-m r}dm\right)
\end{eqnarray}
where $r=|\vec{x}_2-\vec{x}_1|$. Thus, the zero mode is related with the standard electrostatic potential while the massive modes determinate the order of corrections. Similar to the gravitational case,  the heavy modes contribution is exponentially attenuated and  it is expected that the deviation is  generated by the light modes, those with a mass below a critical value, say $m_c$. (The integral in (\ref{Vr}) can be divided into two  integrals: the first one from zero to $m_c$, the second one from $m_c$ to infinity.)  

To find the eigenstates spectrum of (\ref{QQ+psi}) for a domain wall background given by (\ref{Metric}, \ref{scalarfield}, \ref{selfpotential}) is a difficult problem. However, for regularized versions of the RS scenario, far away from the wall, where the effects of the thickness are negligible,  the metric factor is similar to the RS geometry, namely $a(z)\sim-\ln\left(1+\alpha|z|\right)$; and the problem should be simplified. Indeed, for large $z$, the quantum mechanic potential associated to (\ref{QQ+psi})  converges to
\ini\label{VQMAsint}
V_{QM}(z)\sim \frac{\kappa(\kappa+1)}{z^2};
\fin
and the light states density on the wall can be determined as  
\begin{equation}
\psi_m(0)\sim \frac{1}{z_r}\left(\frac{m}{\alpha}\right)^{\kappa-1}.
\end{equation}
See \cite{Csaki:2000fc}, Sec. 4.1 for details. Thus, for $m_c r\gg 1$, the corrections are determined by
\begin{eqnarray}\label{Vr2}
{\mathscr U}(r)\sim\frac{{\cal Q}^2|\psi_0(0)|^2}{4(\sqrt{2\pi})^5}\frac{q_1q_2}{r}\left(1+\frac{\alpha\Gamma(2\kappa-1)}{\pi|\psi_0(0)|^2}\frac{1}{(\alpha r)^{2\kappa-1}}\right).
\end{eqnarray}

\section{Conclusions and comments }

In order to include gauge symmetry in the  localization mechanics (\ref{LA1}),  interaction terms between the vectors and the gravitational field, as indicated in (\ref{Full}), must be considered.  Despite how little conventional of the five-dimensional model, the four-dimensional effective theory obtained after to expand the vectors in the eigenstate basis (\ref{QQ+psi}, \ref{Q+Qpsi}) turned out to be  more conventional: a Maxwell term associated to the massless state of the spectrum and a tower of Stueckelberg fields in correspondence with the massive states, see (\ref{L4}). As a result, $1/r^{2\kappa}$, with $\kappa$ the vector coupling parameter,  was determined as the contribution of the Stueckelberg fields to the four-dimensional electrostatic potential; which, in accordance with (\ref{Vr2}), is negligible for large distance.  

On the other hand, when the action (\ref{LA1}) is considered in the infinitely  thin wall limit, four- and five- dimensional massive term for the vector fields, similar to the Ghoroku-Nakamaura mechanics \cite{Ghoroku:2001zu}, is obtained; which can be identified, respectively, with the tension of the brane and the bulk cosmological constant. Thus, (\ref{LA1})  is a generalization to self-gravitating domain walls of theory (\ref{LGN}).

In a next work we hope to present a covariant generalization of (\ref{Full}). There exist proposals to confine gauge field by means of covariant mechanics;  but, the inclusion of new fields have been required to achieve this end. For example, in \cite{Kehagias:2000au}  the coupling of the vector fields with the dilaton $\pi$ has been considered
\begin{equation}\label{greek}
\frac{{\mathscr L}_A}{\sqrt{-g}}=-\frac{1}{4}e^{2\left(\kappa-\frac{1}{2}\right)\pi}F_{ab}F^{ab},
\end{equation} 
where $\pi$ satisfy the equation of motions of the Bloch brane system
\begin{equation}
\frac{{\mathscr L}_g}{\sqrt{-g}}=\frac{1}{2}R-\frac{1}{2} \nabla^a\phi\nabla_a\phi-\frac{1}{2} \nabla^a\pi\nabla_a\pi-V(\phi,\pi).
\end{equation}
In particular, in conformal coordinates (\ref{Metric}), for $A_b\rightarrow e^{-\left(\kappa-\frac{1}{2}\right)\pi}A_b$ and $\pi\propto a$, we know that (\ref{greek}) converges to (\ref{SAInv}). However, to us we would like to build a covariant version of (\ref{Full}) without having to appeal to new fields that modify the geometry set up by the domain wall.


This work was supported by CDCHT-UCLA under project 008-CT-2012.



\end{document}